\documentstyle[epsf]{article}

\begin{document}
\title{\bf Overview of open issues in the physics of \\
           large solar flares}

\author{
B.\,V. SOMOV\footnote{Corresponding author.
                      Email: somov@sai.msu.ru}\,\,$^{\,1}$,
S.\,I. BEZRODNYKH$^{\,1,\,2}$,
L.\,S. LEDENTSOV$^{\,1,\,3}$,
\\
$^{1}$\,{\small P.K. Sternberg Astronomical Institute,
        Moscow State University,} \\
        {\small Universitetskii Prospekt 13,
        Moscow 119991, Russia} \\
$^{2}$\,{\small A.A. Dorodnitsyn Computational Center,
        Russian Academy of Sciences,} \\
        {\small ul. Vavilova 40, Moscow 119991, Russia} \\
$^{3}$\,{\small Physical Faculty of Moscow State University,} \\
         {\small Leninskie Gori 1, Moscow 119991, Russia} \
        }
  \maketitle

\begin{center}
\parbox{112mm}{ 
\small
A broad variety of observational methods allows us to see the
effect of magnetic reconnection in high-temperature
strongly-magnetized plasma of the solar corona.
Some specific features of the large-scale reconnection in large
solar flares are summarized in this review but they are not
investigated in detail yet.
For example, an analysis of the topological peculiarities of
magnetic field in active regions clearly shows that the so-called
topological trigger phenomenon is necessary to allow for in order
to construct realistic models for large solar flares and
Coronal Mass Ejections (CMEs).
However this is not a simple task.
We discuss also some new analytical models of magnetic reconnection
in a current layer with attached MHD discontinuities.
These models take into account the possibility of a
current layer rupture in the region of anomalous plasma
resistivity.
In the context of the numerical simulations on reconnection,
a question on their interpretation is considered.
Some new results obtained recently are briefly reviewed together
with new questions of the solar flare physics
to be studied.
} 
\end{center}

\vspace{0.5mm}

\noindent
{\em Keywords\/}:
Sun; Solar flares; Magnetic reconnection; Shock waves

%
%
\section{Introduction}
   \label{sec:I}

Large solar flares strongly influence the interplanetary and
terrestrial space by virtue of shock waves, hard electromagnetic
radiation and high-energy accelerated particles
\cite{han07,lil07}.
Early studies of solar flares showed that flares were associated with
magnetic fields
\cite{sev64}.
Estimates of the energy required to power large flares led to the
conclusion that flares must be electromagnetic in origin.
However even much earlier it became more clear, step by step, that a
solar flare is a result of the effect of reconnection of magnetic
fields, the {\em magnetic reconnection\/}
\cite{gio48,swe58,syr62}.
%
%

The principal flare process is contingent on the accumulation of the
{\it free magnetic energy\/} in the corona.
By `free' we mean the surplus energy above that of a potential
magnetic field.
This field has the sources (sunspots, background
fields) in the photosphere.
The free energy is related to electric currents in the corona.
A flare corresponds to rapid changes of the currents.
It is of principal importance to distinguish the currents of
different origin because they have different physical properties and,
as a consequence, different behaviors in the pre-flare and flare
processes (see Sec. 14.5 in \cite{som06}).
The actual currents conventionally comprise two different types.
(a)
The {\em smoothly-distributed\/} currents that are necessarily
parallel or nearly parallel to the field lines, since magnetic field
is strong in the solar atmosphere.
So, the magnetic field is locally force-free (FFF, \cite{aly91}).\,
(b)
The {\em strongly-concentrated\/} electric currents like a
reconnecting current layer (RCL, \cite{syr66}).

It was a question whether or not it is possible to explain the
pre-flare energy storage in a FFF.
However the smoothly-distributed currents dissipate too slowly in
a low-resistivity plasma of the solar atmosphere.
Hence the highly-concentrated currents are necessary to
explain an extremely high power of energy release in the
impulsive phase of a flare.
The RCLs allow an active region to overcome
this difficulty.

We distinguish between two processes: a slow accumulation of energy
and its fast release, a flare.
An interaction of magnetic fluxes (and an excess of energy)
appears as a result of slow changes of the field sources.
The changes are an emergence of a new flux from
below the photosphere and other flows of photospheric plasma, in
particular the {\em shear\/} flows
along the neutral line of the photospheric magnetic field.
So, an actual reconnection in the corona is always a 3D phenomenon
(Sec.~\ref{sec:Tdri}).
%
%
Accumulation of energy in the form of magnetic field of RCLs and
rapid dissipation of the field necessary for a flare can be explained
by the theory of super-hot turbulent-current layers (SHTCL,
\cite{som06,som92})

In general, the potential field determines a large-scale structure
of the flare-active regions while the RCL at separators together
with the other non-potential components of field determine
energetics and dynamics of a large eruptive flare.
To understand the relative role of different currents,
it is necessary to study the evolution of its magnetic
structure in and above the photosphere.
This would allow us to determine not only the magnetic fluxes of
certain magnetic links but also their changes -- redistribution
and reconnection.
Such a study would also give us an information
about the structure and evolution of the electric
field in active regions.
In this article, some new theoretical results obtained recently
are briefly reviewed together with new questions of the solar flare
physics to be studied.

%
%
\section{The topological models of 3D reconnection}
   \label{sec:Tdri}

\begin{figure}[thb]
   \epsfxsize=103mm
   \centerline{\epsfbox{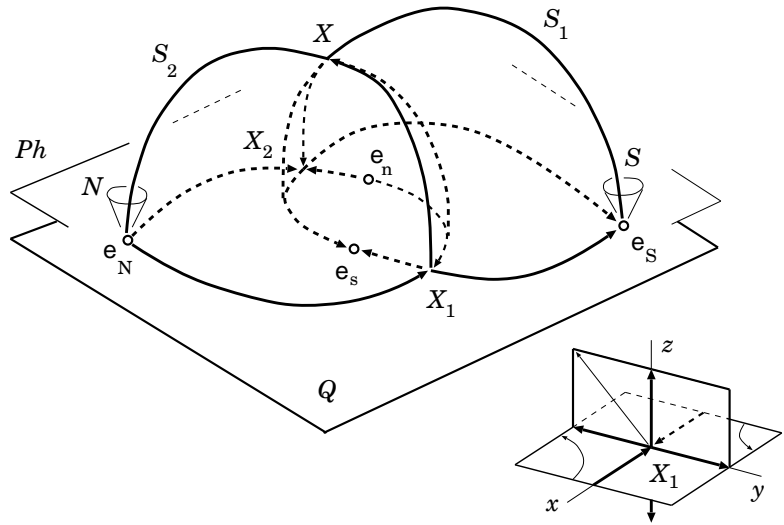}}
\caption{The simplest topological model for magnetic field of four
         sunspots of pairwise opposite polarity.}
   \label{fig1}
\end{figure}

In the simple topological model
\cite{gor90},
four field sources -- the magnetic
``charges''~$ e_{_{\rm N}} $ and $ e_{_{\rm S}} $, $ e_{\rm n} $
and~$ e_{\rm s} $, located in the plane~$ Q $ under the
photosphere~$ Ph $
%
%
(Fig.~\ref{fig1}) -- are used to reproduce the main features of
the observed field in the photosphere related to the four most
important sunspots:~$ N, \, S, \, n $ and~$ s $.
As a consequence, the {\em quad\-ru\-pole\/} model reproduces only the
large-scale features of the actual field in the corona related to
these sunspots.
As a minimum, the four sources are necessary to describe two
interacting magnetic fluxes having the two sources per each.
The larger number of sources are not necessarily much better.

%
%

The main features are two magnetic surfaces called the
{\em se\-pa\-rat\-rices\/}:~$ S_{\rm 1} $ and~$ S_{\, \rm 2} $
(Fig.~\ref{fig1}).
%
%
They divide the whole space above the plane~$ Q $ into four regions
and, cor\-res\-pond\-ing\-ly, the whole field into four magnetic
fluxes having different linkages.
The field lines are grouped into four regions according to their
termini.
The se\-pa\-rat\-rices are formed from lines beginning or
ending at magnetic zeroth points~$ X_{1} $ and~$ X_{\, 2} $.
For example, the field lines originating at the point~$ X_{1} $ form
a se\-pa\-rat\-rix surface~$ S_{\rm 1} $.
%
%

The topologically singular field
line~$ X_{\rm 1} X X_{\rm 2} $, lying at the intersection of the
se\-pa\-rat\-rices, belongs to all four fluxes (the two reconnecting
and two reconnected fluxes) that interact at this line, the 3D
magnetic {\em separator\/}.
So the separator separates the interacting fluxes by
the se\-pa\-rat\-ri\-ces.
Such situation is of fundamental importance for solar physics.
On the other hand, direct detection of a separator, as a field line
that connects two zeroth points in the Earth magne\-to\-tail, by the
four {\em Cluster\/} spacecraft
\cite{xia07}
provides an important step towards establishing an observational
framework of 3D reconnection.

%
%
\section{What is topological trigger?}
   \label{Witt}

Near a separator the longitudinal
component~$ {B}_{\, \parallel} $ dominates because the orthogonal
field~$ {\bf B}_{\bot} $ vanishes at the separator.
Reconnection in the RCL at the separator just conserves the flux of
the longitudinal field,
see Sec. 6.2 in \cite{som06}.
At the separator, the orthogonal components are reconnected.
Therefore they actively participate in the connectivity change,
however the longitudinal field does not.
Thus the longitudinal field plays a passive role in the
topological aspect of the 3D re\-con\-nec\-tion process but it
influences the physical properties of the RCL, in particular the
reconnection rate.
The longitudinal field decreases compressibility of plasma flowing
into the RCL.
When the longitudinal field vanishes at the separator, the plasma
becomes ``strongly compressible'', and the RCL collapses, i.e. its
width decreases substantially.
As a result, the reconnection rate increases quickly.
However this is not the whole story.

\begin{figure} [thb]
\epsfysize=60mm
   \centerline{\epsfbox{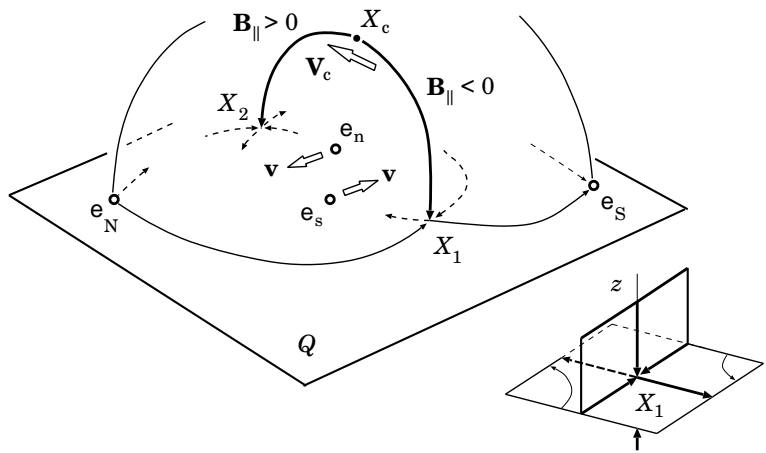}}
   \caption{The zeroth point $ X_{\, \rm c} $ rapidly moves
            along the separator and switches back the longitudinal
            component~$ {\bf B}_{\parallel} $ of magnetic field. }
   \label{fig2}
\end{figure}

The important exception constitutes a zeroth point which can appear
on the separator above the photosphere.
In this case, even very slow changes in the configuration of
field sources in the photosphere can lead to a rapid migration of such
a point along the separator
(Fig.~\ref{fig2})
%
%
%
%
and to a {\em topological trigger\/} of a flare
\cite{gor88}.
This essentially 3D effect is considered in mathematical details in
\cite{som08a,som08b}.

Let us arbitrary fix the positions of three charges, while we move
the fourth one along an arbitrary trajectory in the
plane~$ Q $.
Let an initial position of the moving charge corresponds to the
zeroth points~$ X_{1} $ (with the eigenvalue~$ \lambda_{z}> 0 $)
and~$ X_{\,2} $ (with $ \lambda_{z} < 0 $) respectively
(Fig.~\ref{fig1}).
%
%
The separator is the field line connecting these points without a
coronal null.
This field line emerges from the point~$ X_{1} $ and is directed
along the separator to the point~$ X_{\, 2} $.
%
%

\begin{figure} [tbh]
\epsfysize=67mm
   \centerline{\epsfbox{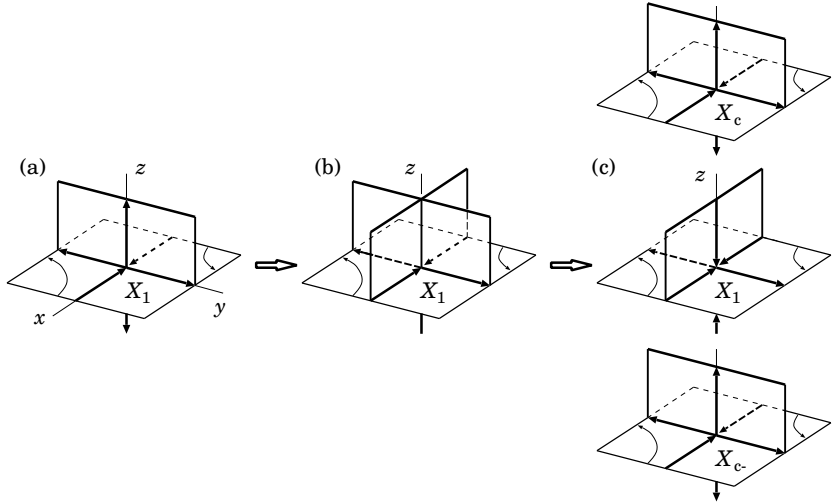}}
   \caption{Changes of the field pattern at the zeroth
            point~$ X_{1\,} $.
   (a) An initial state is the non\-de\-ge\-ne\-rate point
       with $ \lambda_{z}> 0 $.
   (b) A {\em degenerate\/} hyperbolic point (line) with
       $ \lambda_{z}= 0 $
       at the beginning of trigger.
   (c) After the beginning of trigger, the pattern of field is
       the non\-de\-ge\-ne\-rate point
       with $ \lambda_{z}< 0 $ and two zeroth points outside
       the plane~$ z = 0 $. }
   \label{fig3}
\end{figure}

The moving charge can arrive in a narrow region
(let us call it the region~$ TT $) such that both
points in the plane~$ Q $ will have the same sign of~$ \lambda_{z} $
(Fig.~\ref{fig2}).
%
%
In this case there must also exist two zeroth points outside the plane.
They are arranged sym\-met\-ri\-cally relative to the plane~$ Q $ (the
plane~$ z = 0 $).
Figure~\ref{fig3}
%
%
illustrates how these additional points appear.
%
%
%
Before the start of trigger, the moving charge is outside of the
region~$ TT $, and $ \lambda_{z}> 0 $ at the non\-de\-ge\-ne\-rate
zeroth point~$ X_{1} $
(Fig.~\ref{fig3}a).
%
%
When the moving charge crosses the boundary of the region~$ TT $,
the eigenvalue~$\lambda_{z}$ at the point~$ X_{1} $ vanishes
(Fig.~\ref{fig3}b):
%
%
\begin{equation}
   \lambda_{z}( X_{1} ) = 0 \, .
    \label{080101}
\end{equation}
The point becomes degenerate.
At this instant, another pair of zeroth points is born from
the point~$ X_{1} $
(Fig.~\ref{fig3}c).
%
%
Consider one of them, the point~$ X_{c} $ in the upper
half-space~$ z > 0 $.
This  non\-de\-ge\-ne\-rate point travels along the separator and
merges with the point~$ X_{\, 2} $ in the plane~$ z = 0 $ when the
moving charge emerges from the region~$ TT $.
At this instant $ \lambda_{z}$
vanishes:
\begin{equation}
   \lambda_{z}( X_{\, 2} ) = 0 \, .
    \label{080102}
\end{equation}
As a result of the process described, the direction of the field at
the separator is reversed with the point~$ X_{\,2} $ with
$\lambda_{z}> 0 $.
After that, the moving charge is located outside the region~$ TT $,
there are no zeroth points outside the plane~$ z = 0 $.
Thus Equations~(\ref{080101}) and~(\ref{080102}) determine the
boundaries of the topological trigger region~$ TT $.
%
%

\begin{figure}
\epsfysize=140mm
   \centerline{\epsfbox{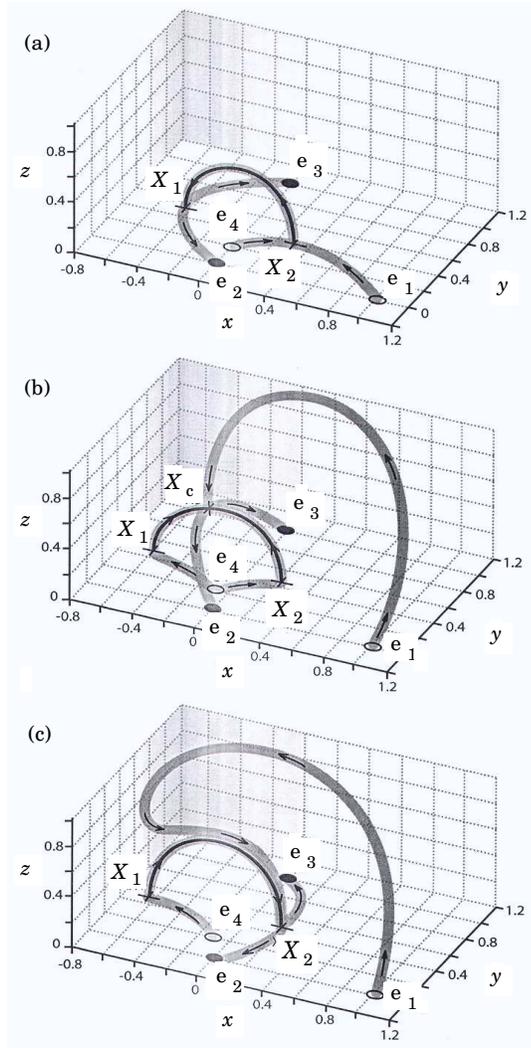}}
   \caption{Drastic changes of the large-scale structure of
       magnetic field in a model of active region.
   (a) An initial state without a coronal null of magnetic field.
   (b) The coronal null~$ X_{c}$ runs along the separator during the
       topological trigger process.
   (c) After the end of the trigger process. }
   \label{fig4}
\end{figure}

Typically the region~$ TT $ is very narrow.
That is why small shifts of the moving charge within this region lead
to large shifts of the zeroth point~$ X_{c} $ along the separator
above the plane~$ z = 0 $ just creating a {\em global bifurcation\/}.
By analogy with hyd\-ro\-dy\-na\-mics
\cite{ore-i-09},
the se\-pa\-rat\-rix
plane~($ y, z $) in
Fig.~\ref{fig3}a,
%
%
which plays the role of a `hard wall' for
`flowing in' magnetic flux, is quickly replaced by the orthogonal
hard wall, the se\-pa\-rat\-rix
plane~($ x, z $) in
Fig.~\ref{fig3}c.
%
%
Thus the topological trigger drastically changes directions of
magnetic fluxes in an AR as illustrated by
Fig.~\ref{fig4}.
%
%
%
%
During the quick process of topological trigger, another rigid wall,
the se\-pa\-rat\-rix plane of the coronal zeroth point~$ X_{c} $
running along the separator, turns the large-scale magnetic field of
an active region like a page in a book.

Figure~\ref{fig4}b
%
%
shows that a loop~$ e_{1} X_{c} $ quickly grows
up.
If it reaches a height in the corona, where the solar wind becomes
important and pools the magnetic field lines in the interplanetary
space
\cite{som72},
then a fast motion appears as an upward collimated jet along a
coronal streamer structure.
%
%

We have considered the travel of one charge while the
coordinates and magnitudes of the other three charges were fixed.
It is obvious however that all the foregoing remains in force in
a more general case of variation of the charge configuration.
A slow evolution of the configuration
of field sources in the photosphere can lead to a rapid
rearrangement of the global topology in active regions in
the corona.
So the phenomenon of topological trigger is necessary to
model the large eruptive flares.

Note that the topological trigger effect is {\em not\/} a resistive
instability which leads to a change of the topology of the field
configuration from pre- to post reconnection state.
On the contrary, the topological trigger is a quick change of the
global topology, which dictates the fast reconnection of col\-lisional
or col\-li\-sion\-less origin.
Thus the term `topological trigger' is the most appropriate
nomenclature to
emphasize the basic nature of the topological effect involved, and
it is a welcome usage.
%
%

In close relation to the effect of topological trigger, we have to
undertake a series of extensive researches to elucidate the
electrodynamic processes in the photosphere, chromosphere and corona,
related to magnetic field concentration and dissipation,
plasma heating and acceleration of electrons and ions.
A major thrust of this work is the development of a unified model
for electrodynamic coupling at all levels in the solar atmosphere
starting from the photosphere.
This is what we need to involve the topological trigger in a
mechanism of the geo\-ef\-fec\-tive solar events and to develop
their physics.
Such investigations will also give us the possibility of applying
the theory of large solar-type flares to astrophysical phenomena
(like the flares in accretion disk coronae) accompanied by fast
plasma ejections, powerful fluxes of heat and radiation, impulsive
acceleration of electrons and ions to high energies
\cite{som03}.

%
%
\section{New analytical models of reconnection} 
   \label{sec:Nam}

\begin{figure} [thb]
\epsfysize=100mm
   \centerline{\epsfbox{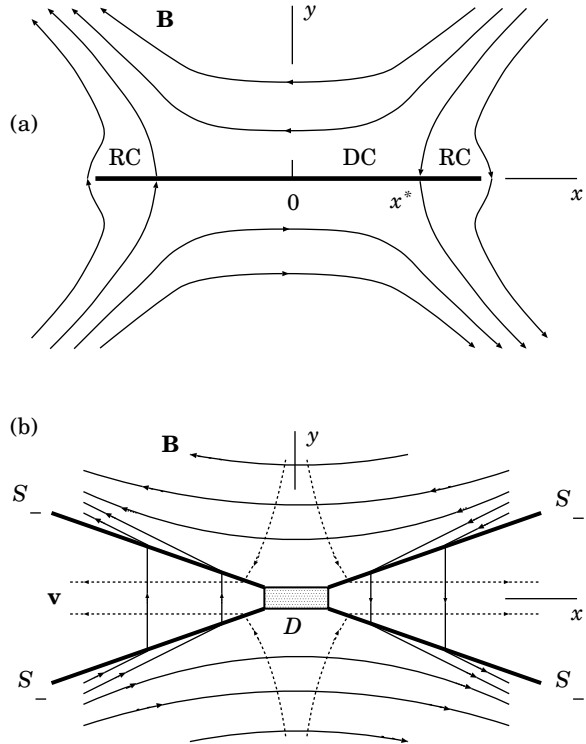}}
   \caption{Two classical models of reconnection.
   (a) Sy\-ro\-vat\-skii's current layer contains a region of direct
       current (DC) and two regions of reverse current (RC).
   (b) Pet\-schek's flow consists of a small diffusive region~$ D $ and
       four attached slow MHD shocks~$ S_{-} $. }
   \label{fig5}
\end{figure}

In the approximation of a strong magnetic field,
Sy\-ro\-vat\-skii \cite{syr71}
constructed a simple 2D analytical RCL model in the form of a
discontinuity surface that separates the oppositely directed magnetic
fluxes
(Fig.~\ref{fig5}a).
%
%
%
%

Another classical model
(Fig.~\ref{fig5}b)
%
%
is called Pet\-schek's flow
\cite{pet64}
and is commonly considered as an alternative to Sy\-ro\-vat\-skii's
current layer.
Mar\-kovskii and So\-mov~\cite{mar89}
suggested a stationary reconnection model that is a generalization of
both models.
This model consists of an infinitely thin current layer of length~$2a$
with four MHD shock waves of length~$ r $ attached to its endpoints
like whiskers.
The normal magnetic field component vanishes on the current layer and
is equal to a given constant~$ \beta $ on the shock waves inclined to
the current layer at angle~$ \alpha $.

\begin{figure} 
   \epsfysize=76mm
      \vspace{2mm}
      \centerline{\epsfbox{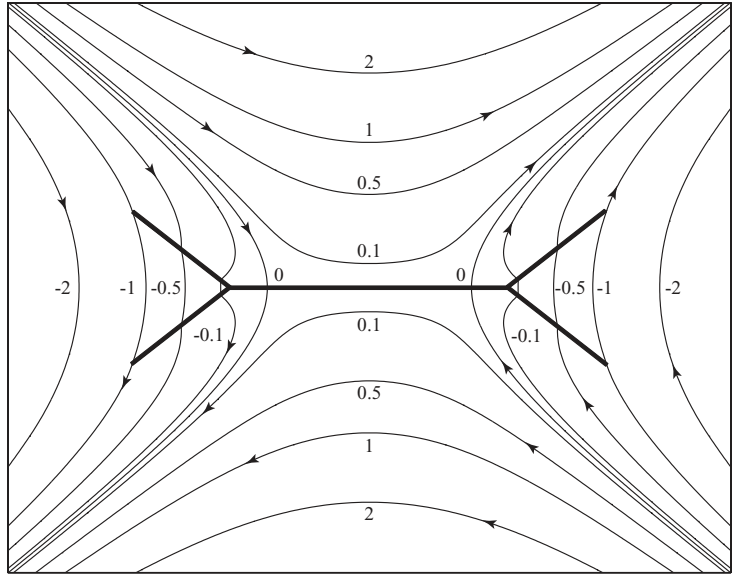}}
      \vspace{2mm}
   \caption{Magnetic field lines in the pattern which is typical for
            general case of physically meaningful solutions of
            the reconnection problem. }
   \label{fig6}
\end{figure}

An asymp\-to\-tics of the solution to this problem
corresponding to a small whisker length~$ r $ was found in
\cite{mar89}.
We present the general solution of the problem with
ar\-bit\-rary length~$ r $ as illustrated by
Fig.~\ref{fig6}
taken from \cite{bez11}.
%
%
%
%
This solution is found in an analytical form
that admits of efficient numerical implementation.
So we can analyze in detail the structure of the magnetic field and
its variation with reconnection-model parameters.

In the center part of the pattern we see the thin-wide current layer
with the region of direct current and two regions of reverse current
in a good agreement with Sy\-ro\-vat\-skii's model.
In addition, four MHD discontinuities are attached to the edges of the
layer.
Note that a character of discontinuous flows is not prescribed but
determined from a self-consistent solution of the problem
in the approximation of strong field.
Should the discontinuous flows be similar to the slow MHD shock
waves in Pet\-schek's flow? --
No, these discontinuities are not so simple.
The general solution shows that a situation is more
complicated.

In different parts of the attached discontinuities, different types of
discontinuous solutions present.
For example, slow shocks associated with the reconnection process
in Pet\-schek's flow are separated
from the current layer by fast MHD shocks.
Moreover, near the edges of the layer with return currents
inside it, there appears the regions of trans-Alfv\'{e}nic shock
waves (TASW)
\cite{led11}.
The distinctive feature of TASW is that they do not exist as
stationary shocks: they disintegrate, transform to
time-dependent compound waves, or evolve to Alfv\'{e}n waves
in a diffusion-like manner
\cite{mar00,fal01}.
As a consequence, some parts of the attached discontinuities become
non-evolutionary (see Ch.~17 in
\cite{som06-I}).
This fact gives us the principal possibility to understand general
properties of reconnection in the framework of 2D resistive
theory and direct numerical simulations,
e.g., \cite{bis97,buc03}.

\begin{figure} 
   \epsfysize=76mm
      \vspace{2mm}
      \centerline{\epsfbox{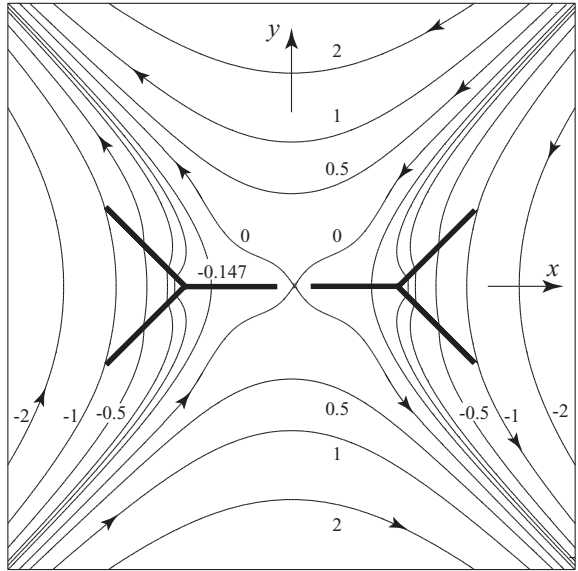}}
      \vspace{2mm}
   \caption{Magnetic field lines in the vicinity of a
            disrupted current layer with attached discontinuous
            MHD flows. }
   \label{fig7}
\end{figure}

The most interesting problem is the behavior of magnetic field and
plasma in the edge region of a RCL, i.e. in a complicated transition
from the RCL to the exterior region of a `downstream cone'
between two attached MHD shocks.
Another important problem is to find the generalized analytical
solutions those take into
account the possibility of the current-layer rupture
(Fig.~\ref{fig7})
%
%
%
%
in a place of anomalous resistivity of plasma
\cite{bez11}.
The current layer rupture region is assumed to be the place of
particle acceleration to high energies in solar flares
\cite{som10a}.
The regions of direct and reverse currents and the refraction of
magnetic field lines at the attached shocks are clearly seen in
Fig.~\ref{fig7}.
%
%
Thus we can study the global structure of magnetic field in a
reconnection region as well as the local properties of the field
in the vicinity of the current layer and attached discontinuities.
%

%
%
\section{New problems in the solar flare physics} 
   \label{sec:OIRF}

The existing topological models are used to describe active
regions, to reproduce the main features
of magnetic fields leading to the large-scale reconnection in solar
flares
\cite{som02,som10b}.
To understand the 3D structure of reconnection in flares is
one of the most urgent problems.
{\em RHESSI\/}, {\em Hinode\/} and other modern space missions
offer us the
means to check whether phenomena predicted by topological models
(such as the topological trigger) do occur.
However some puzzling discrepancies may also exist, and further
development of realistic 3D models is required.

The local models of reconnection in flares take kinetic
effects into account and
allow us to develop the basic physics of reconnection in flares
\cite{som06,som10a}.
Col\-li\-sion\-less reconnection is a key process in flares.
It was introduced by
Sy\-ro\-vatskii
\cite{syr66}
as a {\em dynamic dissipation\/} of field in a current layer and leads
to fast conversion from field energy to particle energy.
The problem of stable motion of particles in a RCL was considered
in the adiabatic approximation with account of three components of
magnetic field and an inductive electric field related to
reconnection
\cite{ore-a-09}.
However the non-adiabatic behavior of accelerated particles in the
RCL is a big issue here.

General properties and parameters of the col\-li\-sion\-less
reconnection can be examined in a frame of models based on the mass,
momentum, and energy conservation laws.
A particular feature of the models is that electrons and ions are
heated by wave-particle interactions in a different way.
The magnetic-field-aligned thermal flux becomes anomalous and plays
the role in the cooling of electrons in a SHTCL
\cite{som06}.
These properties are typical for the conditions derived
from the observations by {\em RHESSI\/}
\cite{liu08}.
Unfortunately, the local models are not incorporated in the global
consideration of reconnection in the corona.
Only a few first steps have been made in this direction.
%
%

Modern spacecraft observations of col\-li\-sion\-less reconnection
in the magne\-to\-tail and magne\-to\-pause as well as recent
simulations show the existence of {\em thin\/} current layers with
scale lengths of the order a few electron skin depth,
e.g., \cite{zei02}.
In the electron MHD model of such current layers, reconnection is
facilitated by electron inertia which breaks the frozen-in condition.
The simulations demonstrate the instability of the whistler-like
mode which presumably plays the role of the ion-cyclotron instability
in the SHTCL model.

Future models of flares should join global and local properties
of reconnection under coronal conditions.
For example, chains of plasma instabilities, including kinetic
instabilities, can be important for our understanding the types
and regimes of plasma turbulence inside the col\-li\-sion\-less
current layers with a longitudinal magnetic field.
In particular it is necessary to evaluate better the anomalous
resistivity and selective heating of particles in such a SHTCL.
Heat conduction is also anomalous in the super-hot plasma.
However the effect of col\-li\-sional relaxation of heat fluxes
(e.g., \cite{gol77})
from SHTCL should be taken into account and presumably better describes
heat transfer in flares than the classic Fourier law and the
anomalous heat conduction
\cite{ore-a-11}.

Self-consistent solutions of the reconnection problem will allow us to
explain the energy release in flares, including the open question of
the mechanism or combination of mechanisms which explains the observed
acceleration of electrons and ions to high energy.
The problem of particle acceleration in the collapsing magnetic traps
created by reconnection in the corona
\cite{som97}
is considered by taking into
account the particle scattering and braking in the high-temperature
plasma of solar flares
\cite{bog09}.
The Coulomb collisions are shown to be weak in traps with life-times
$ \tau < 10 $~s and strong for~$ \tau > 100 $~s.
For col\-li\-sional times comparable to~$ \tau $, the electron spectrum
at energy above 10-20~keV is shown to be a double-power-law one.
Such spectra in hard X-rays are often found by {\em RHESSI\/} in
flares.

However note that the most sensitive tool to study behavior of the
electron acceleration in the collapsing trap is radio-emission.
At least, the simple gyro\-synchrotron radio-emission
has to be calculated for evolving electron distribution in
a collapsing magnetic trap in order to understand
how the magnetic trap and the particle distribution evolve
\cite{li09}.
Wave-particle interactions can presumably be important too;
see Sec.~7.4.4 in
\cite{som06}.
Resonant scattering is likely to enhance the rate of precipitation of
the trapped electrons with energy higher than 100 keV, generating
microwave bursts.

The lose-cone instabilities of trapped mildly-relativistic electrons
could provide excitation of waves with a very wide continuum spectrum.
In a solar flare with a slowly-moving upward coronal HXR source,
an ensemble of the collapsing field lines with accelerated
electrons would presumably be observed as a slowly moving type IV burst
with a very high brightness temperature and with a possibly significant
time delay relative to the chromospheric foot\-point emission.
The second kind of wave-particle interactions in the collapsing
trap-plus-precipitation model is the streaming instabilities
(including the current instabilities related to a reverse current)
associated with the precipitating electrons.
%
%

The reverse-current electric field results in an essential change of
the fast electron behavior in the thick target.
It leads to a quicker decrease of the fast-electron number with a
column depth in comparison with the classical thick-target model.
It makes the fast-electron distribution to be more isotropic and
leads to a significant decrease of expected hard X-ray bremsstrahlung
polarization; see Sec.~4.5 in
\cite{som06-I}.
Future models should incorporate such fine (or even subtle) effects
like an initial nonuniform ionization of chromospheric plasma in
the thick target,
the time-of-flight effect etc. with correct account taken of the
reverse-current electric field as the effect of primary importance.
Otherwise the accuracy of the existing models will be lower that
the accuracy of new observations of solar flares in hard X-rays.

%
%
\section*{Acknowledgements}

This work was supported by the Russian Foundation for Fundamental
Research (project No.~0802-01033-a).

%

%
%


\end{document}